\documentclass[letterpaper,twocolumn,10pt]{article}
\usepackage{usenix2019_v3}

\usepackage{cite}
\usepackage{courier}
\usepackage{amsmath,amssymb,amsfonts}
\usepackage{graphicx}
\usepackage{textcomp}
\usepackage{xcolor}
\usepackage{tikz}
\usepackage{amsmath}
\usepackage{tabularx}
\usepackage{booktabs}
\usepackage{xurl}
\usepackage{hyperref}
\usepackage{listings}
\usepackage{listings}
\usepackage{lstautogobble} 
\usepackage{color} 
\definecolor{bluekeywords}{rgb}{0.13, 0.13, 1}
\definecolor{greencomments}{rgb}{0, 0.5, 0}
\definecolor{redstrings}{rgb}{0.9, 0, 0}
\definecolor{graynumbers}{rgb}{0.5, 0.5, 0.5}
\usepackage{threeparttable}

\usepackage{authblk}
\usepackage{algorithm}
\usepackage{algpseudocode}
\usepackage{amsmath}

\DeclareUnicodeCharacter{FB01}{fi}
\hypersetup{
    colorlinks=true,
    linkcolor=black,
    filecolor=black,
    urlcolor=black,
    citecolor=black,
}
\usepackage{ragged2e}
\usepackage{verbatim}
\def\BibTeX{{\rm B\kern-.05em{\sc i\kern-.025em b}\kern-.08em
    T\kern-.1667em\lower.7ex\hbox{E}\kern-.125emX}}
\begin{document}

\title{\Large Firmware Re-hosting 
Through Static Binary-level Porting}

\author[1]{Mingfeng Xin}
\author[1]{Hui Wen}
\author[1]{Liting Deng}
\author[1]{Hong Li}
\author[2]{Qiang Li}
\author[1]{Limin Sun}

\affil[1]{Institute of Information Engineering, CAS, China}

\affil[2]{School of Computer and Information Technology, Beijing Jiaotong University, China}

\affil[ ]{\textit{\{xinmingfeng,wenhui,dengliting,lihong,sunlimin\}@iie.ac.cn}, liqiang@bjtu.edu.cn}

\maketitle

\begin{abstract}
The rapid growth of the Industrial Internet of Things (IIoT) has brought embedded systems into focus as major targets for both security analysts and malicious adversaries. Due to the non-standard hardware and diverse software, embedded devices present unique challenges to security analysts for the accurate analysis of firmware binaries. The diversity in hardware components and tight coupling between firmware and hardware makes it hard to perform dynamic analysis, which must have the ability to execute firmware code in virtualized environments. However, emulating the large expanse of hardware peripherals makes analysts have to frequently modify the emulator for executing various firmware code in different virtualized environments, greatly limiting the ability of security analysis.

In this work, we explore the problem of firmware re-hosting related to the real-time operating system (RTOS). Specifically, developers create a Board Support Package (BSP) and develop device drivers to make that RTOS run on their platform. By providing high-level replacements for BSP routines and device drivers, we can make the minimal modification of the firmware that to be migrated from its original hardware environment into a virtualized one. We show that an approach capable of offering the ability to execute firmware at scale through the use of firmware patching in an automated manner without modifying the existing emulators. Our approach, called \textit{static binary-level porting}, first identifies the BSP and device drivers in target firmware, then patches the firmware with pre-built BSP routines and drivers that can be adapted to the existing emulator. Finally, we demonstrate the practicality of the proposed method on multiple hardware platforms and firmware samples for security analysis. The result shows that the approach is flexible enough to emulate firmware for vulnerability assessment and exploits development.

\end{abstract}

\section{Introduction}
With the proliferation of Industrial Internet of Things (IIoT), embedded devices, such as routers and many control devices, are continuously increasing and used in many industrial application domains. These devices are usually connected to the different types of networks for extra functionality, executing a special-purpose computing task with co-operations. However, the connectivity of embedded devices significantly increased their exposure to attacks. It makes analysts focus on the security analysis of embedded device firmware, aiming to discover vulnerabilities for system security assessment and firmware patching.
Costin et al.\cite{CostinLargeScale} perform a large-scale analysis of the security of embedded firmware through static analysis. 
However, dynamic analysis plays a crucial role when researchers want to conduct a thorough security analysis on a specific firmware.
As far as is known, dynamic analysis can perform a wide range of sophisticated examinations (e.g., taint analysis and symbolic execution) and overcome the limitations of static analysis (e.g., packed or obfuscated code). In terms of firmware analysis, dynamic analysis of embedded systems can monitor detailed data flow, including memory layout, register values, etc., greatly improving the analysis ability of vulnerabilities of embedded devices.

Unfortunately, embedded hardware provides limited introspection capabilities, including limited numbers of breakpoints and watchpoints, significantly restricting the ability of dynamic analysis on firmware. In this case, emulation, also known as firmware re-hosting, enables the host system to execute the firmware of embedded hardware in virtualized environments for successfully performing dynamic analysis on firmware. However, appropriate emulators are typically rare, particularly due to the impracticality of drivers for supporting incompatible embedded processors. Moreover, the embedded system relies on different peripherals and system configurations. It makes various peripherals and memory layouts that must be supported in a specialized manner by emulators. In conclusion, the heterogeneity in embedded hardware makes it hard to decouple the firmware from the hardware and emulate a large number of hardware peripherals.

To solve this problem, researchers propose many emulation solutions for firmware analysis, such as Avatar\cite{Zaddach2014}, Pretender\cite{Gustafson2019}, P$^2$IM\cite{Feng2020} and HALucinator\cite{Clements2020}. They present a considerable performance but expose their own problems when emulating firmware in a specific condition. For example, Avatar is a hardware-dependent, one-to-one dynamic analysis framework, which must use a debug interface to interact with a physical device. This hardware-in-the-loop design greatly limits the scale of firmware re-hosting. Pretender and P$^2$IM model the peripherals to provide proper values for the fuzzer thus ensure sufficient code coverage, but bring uncertainty with unknown execution process. Specifically, an observation shows that the emulated results conducted by their methods present a value that out the range of the real device should have. HALucinator utilizes High-Level emulation method that provides simple, analyst-created, and high-level replacements, which performs the same task from the ﬁrmware's perspective. However, it only solves emulation problems for Arm Cortex-M-based firmware and cannot emulate firmware running on x86, MIPS, or other processors. In conclusion, these firmware re-hosting solutions only target a specific set of firmware, such as Arm-based firmware. When analyzing a firmware running on the other processors, researchers have to re-implement the emulator, which requires a huge amount of manual work.

Emulators are a key component in enabling dynamic analysis of the firmware by emulating virtual interactions that matches the real system.
According to WRIGHT et al.'s work\cite{wright2020challenges}, \textit{execution fidelity} describes how closely execution in the emulator can match that of the physical system, while \textit{data} or \textit{memory fidelity} describes in what level the memory in the emulator (e.g., BlackBox, RAM, Register) is consistent with hardware for a given point in execution. When re-hosting firmware for Fuzz testing, the data fidelity can be very low. However, when used for vulnerability assessment, the solution of firmware re-hosting needs high fidelity, because the proof of concepts (PoCs) or exploits developed under an emulated environment have to be applied to the real device.

Nowadays, most of embedded systems use RTOS for multiple tasks management to ensure better program flow and event response. In embedded systems, Board Support Package (BSP) is the layer of software containing hardware-specific routines that allow a particular RTOS to function in a particular hardware environment. The purpose of a BSP is to configure the kernel for the specific hardware on the target board. However, the BSP does not include much more than what is needed to support a minimum number of peripherals on a board, and developers have to create the remaining device drivers following the \textit{Device Driver Infrastructure} defined in RTOS. The Device Driver Infrastructure defines device models and driver models for different kinds of devices. When the RTOS is running, the device driver creates a device instance with hardware access capabilities based on the device model definition and registers the device into the device manager or the kernel, and the application will call an interface function to find the device and access hardware via the driver.
It gives us an intuitive way to implement BSP and driver replacements using the abstraction provided by the RTOS to solve heterogeneity for firmware re-hosting. By providing customized replacements for BSP routines and device drivers, it is easier for developers to port firmware from one board to another.

Based on this observation, we solve the re-hosting problem by treating firmware re-hosting as porting firmware to the emulator's board. We propose a method called \textit{static binary-level porting} and build a tool called FirmPorter, which can modify the BSP and driver code at binary level. The tool automatically replaces the BSP and drivers with proper functions that the emulator can successfully execute in a virtual environment. In this work, the tool first identifies BSP and drivers in the firmware for function replacement. Then it generates object files to patch the firmware following the RTOS’s programming interface. Finally, when dealing with peripherals not supported by the emulator, we use a new technology called \textit{Driver Hacking} to solve the communication problems between the host and the emulator.

\noindent\textbf{Contributions:}

1. We explore the problem of firmware re-hosting related to real-time operating system (RTOS) and show an approach that can customize function replacements for Board Support Package (BSP) and device drivers, which is able to handle diverse firmware.

2. We build FirmPorter, a tool that can emulate RTOS firmware through static binary-level porting technique by providing function replacements for BSP and device drivers.

3. We evaluate 11 different firmware with 4 kinds of RTOSes running on 5 different processors, including a Schneider SAGE2400 RTU firmware, a Schneider Modicon M241 PLC firmware and three wireless temperature sensor firmware, which run on real-world devices. The result shows that firmware can be correctly emulated, and PoCs/exploits developed under the emulated environment can be directly applied to real devices.

\section{Background and Motivation}
\label{background}
In this section, we briefly introduce the Board Support Package (BSP) and Device Driver Infrastructure that exist in modern RTOSes, which inspires us to solve the problem of re-hosting in a novel way.

\subsection{Board Support Package}
The purpose of a BSP is to configure the kernel for the specific hardware on the target board, which is the basis of RTOS porting. There is no official definition of BSP yet. According to Wikipedia, a board support package (BSP) is the layer of software containing hardware-specific drivers and other routines that allow a particular operating system (traditionally a real-time operating system, or RTOS) to function in a particular hardware environment (a computer or CPU card), integrated with the RTOS itself. From the view of developers, the BSP allows for a well-defined interface between the target hardware and the operating system. During the boot process, the BSP routines must call core OS routines and device driver routines to configure a portion of the core OS as well as the device drivers. The OS and well-written device drivers then make calls to the BSP routines during system operation to make specific hardware requests.

The essence of porting a BSP is to implement different routines to provide specific hardware-related operations. For example, in VxWorks, \texttt{sysClkEnable()} is to turn on system clock interrupts, and the \texttt{sysNvRamGet()} is to get the contents of non-volatile RAM. Developers must realize these routines according to the hardware datasheets. Table \ref{tbbsproutines} lists part of the routines in VxWorks BSP.

Due to the well-designed abstraction of the RTOS, we are able to separate the BSP part from the project. In the RTOS boot sequence, we consider the range from the first instruction executed after reset up to the initialization of the RTOS kernel as the BSP part. For example, in VxWorks, we consider the code executed before \texttt{usrRoot} as the BSP part, which includes all the BSP routines.

\begin{figure}[t]
\centering
\includegraphics[scale=0.4]{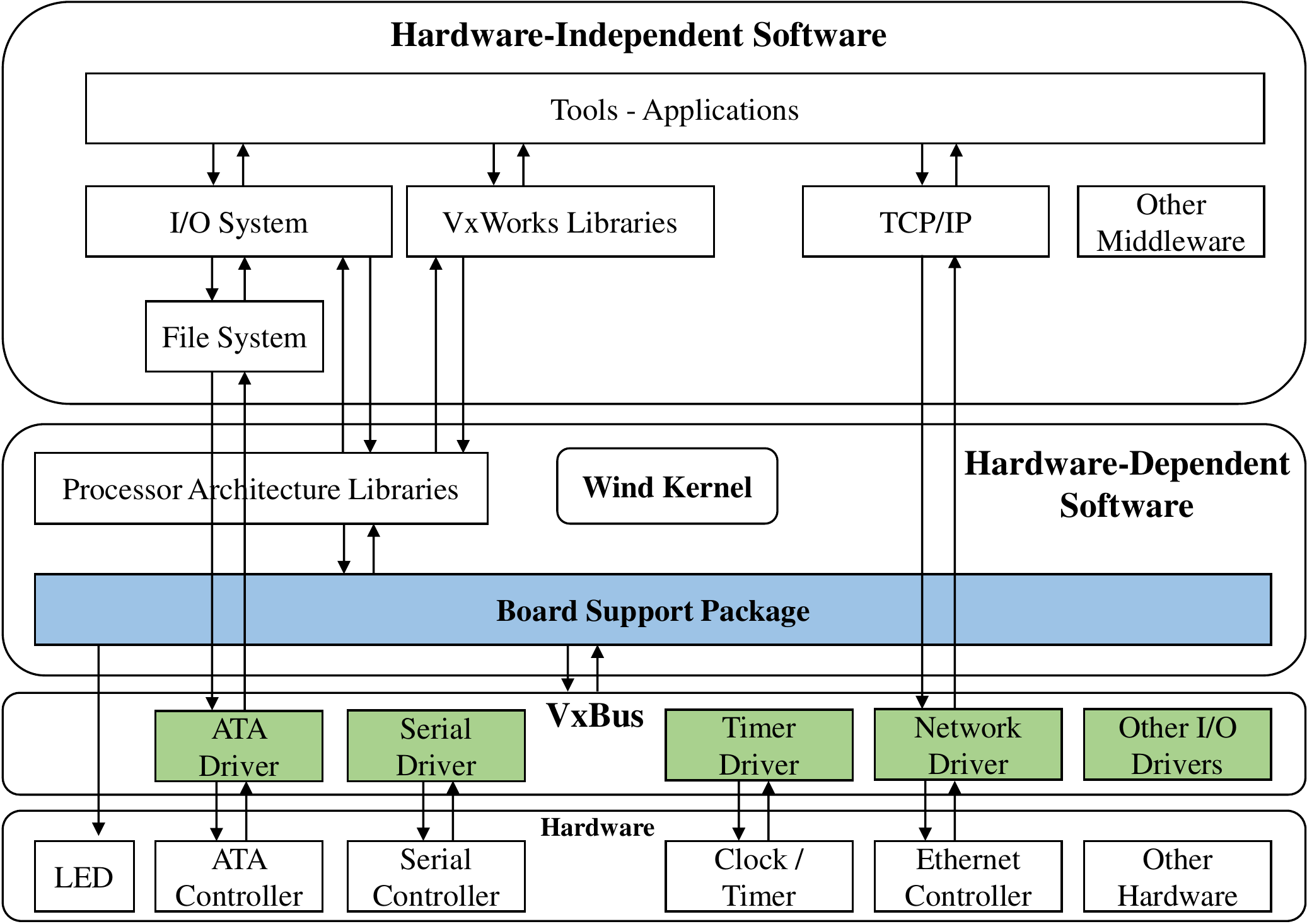}
\caption{Replacing BSP and device drivers when porting a VxWorks-based firmware}
\label{figdrvmod}
\end{figure}

\begin{table}[t]\scriptsize
\centering
\caption{Part of VxWorks BSP routines}
\begin{tabular}{p{2cm}p{5.2cm}}
    \toprule
        Routine &Function\\
    \midrule\midrule
    	\texttt{sysClkInt()} &handels system clock interrupts \\
    	\texttt{sysClkEnable()} &turns on system clock interrupts \\
    	\texttt{sysHwInit()} &initializes the system hardware to a quiescent state \\
    	\texttt{sysHwInit2()} &initializes and configures additional system hardware \\
    	\texttt{sysNvRamGet()} &gets the contents of NVRAM \\
    	\texttt{sysNvRamSet()} &sets the contents of NVRAM \\
    	\texttt{sysToMonitor()} &transfers control to ROM monitor \\
    	$\cdots$ &$\cdots$ \\

    \bottomrule
\end{tabular}
\label{tbbsproutines}
\end{table}
\subsection{Device Driver Infrastructure}
\begin{table}[t]\scriptsize
\centering
\begin{threeparttable}[b]
\caption{RTOS with/without a device driver infrastructure}
\begin{tabular}{p{2cm}<{\centering}p{0.5cm}<{\centering}p{2cm}<{\centering}p{0.5cm}<{\centering}}
    \toprule
        RTOS &Y/N &RTOS &Y/N\\
    \midrule\midrule
    	Zephyr\cite{zephyr} &Y &uC/OS-III &N\\
		VxWorks &Y &TencentOS-tiny &N\\
		RT-Thread\cite{rtthread} &Y &Mbed OS &N\\
		RIOT &Y &FreeRTOS\cite{freertos} &N\\
		Phoenix-RTOS &Y &Contiki-NG &N\\
		Nuttx\cite{nuttx} &Y &ChibiOS &N\\
		Nucleus &Y &Azure RTOS &N\\
		Apache Mynewt &Y &AliOS Things &N\\

    \bottomrule
\end{tabular}
\label{tbrtosdrvmod}
\end{threeparttable}
\end{table}
Device drivers are the liaison between the hardware and the operating system, middleware, and application layers. They are the software libraries that initialize the hardware and manage access to the hardware by higher layers of software. 

In the late 80s, developers access the peripherals by directly manipulating the MMIO for efficiency. For example, a developer wants to read data from the UART, she/he may write the following code:

\lstset{language=C}
\begin{lstlisting}
#define DATA_REGISTER 0xF00000F5
char getchar(){
    return (*((char*)DATA_REGISTER));
\end{lstlisting}

With the idea of decoupling software and hardware prevailing, some well-encapsulated library functions appear. For example, HAL(Hardware Abstraction Layer) libraries such as STM32CubeMX HAL\cite{cubemx} provide a well-packaged function library for Arm's common peripheral functions. In STM32CubeMX HAL, if a developer wants to read the data from the UART, she/he can call the following function:

\lstset{language=C}
\begin{lstlisting}
HAL_StatusTypeDef HAL_UART_Receive( ... ) 
{ ......
  tmp = (uint16_t*)pData;
  *tmp = (uint16_t)(huart->Instance->DR &   (uint16_t)0x01FF); 
  ......
} 

\end{lstlisting}

With the popularity of Linux and the enhancement of embedded device performance, device driver infrastructures have gradually appeared in RTOS. The device driver infrastructure provides a set of standard interfaces, which can reduce the difficulty of driver development. The maintainer of the driver designs a set of mature, standard, and typical driver implementations for each type of driver, and extracts the same parts of the same hardware drivers from different manufacturers to implement them by themselves, and then leaves the different parts out of the interface for the specific engineer implements it. For example, the read operations of all upper-level functions call the \texttt{read()} function, and no longer distinguish between \texttt{uart\_read()} or \texttt{spi\_read()}.

\begin{figure}[t]
\centering
\includegraphics[scale=0.5]{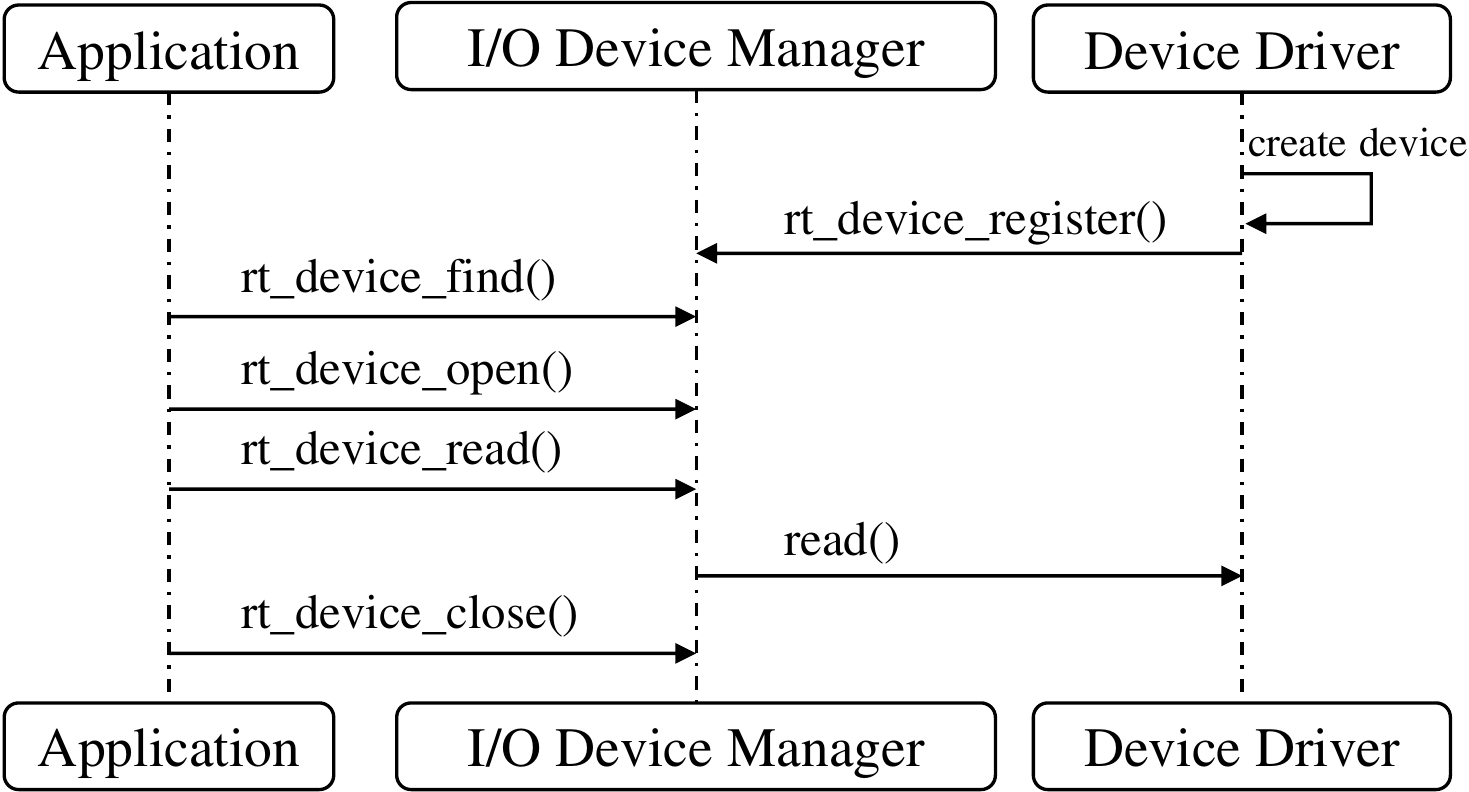}
\caption{Sequence of creating and registering I/O devices in RT-Thread}
\label{figrtdrv}
\end{figure}

A typical example is RT-Thread, which divides peripherals into 14 types and provides different peripheral driver models for different types of peripherals. Drivers are uniformly maintained by the RT-Thread \textit{I/O Management Framework}. When an application needs to call a certain hardware function, it follows the route in Fig.\ref{figrtdrv}.

One can observe that there are some specific interfaces and data structures in the device driver infrastructure. For example, each driver finally appears as a complicated structure, in which the description and functions of the device are defined, such as data reading and writing, interrupt processing, etc. The system registers it in the driver management framework by calling the initialization function, and the application calls specific functions by finding the driver's interface function.

This structure is also reflected at the binary level. For a well-packaged RTOS, the outermost layer of its device driver is usually a wrapper function, which initializes the device structure internally and installs it into the kernel. When the RTOS initializes the device driver, it usually calls an initialization function or uses a function pointer.

\subsection{Motivation}
The performance of firmware re-hosting is limited by the ability of emulation of diverse hardware. Current solutions allowing for the emulation of diverse hardware rely on real devices, hardware models,  or HAL handlers, where the emulator can interact with unsupported peripherals. However, these methods need to modify a set of different emulators for adapting various CPU architectures and libraries. 

Fortunately, most of the embedded devices use RTOS, which contains a Board Support Package (BSP), along with a Device Driver Infrastructure that helps developers customize their hardware drivers.
To avoid the extensive modifications of emulators when re-hosting various firmware, we aim to design a scalable and flexible firmware re-hosting framework to support hardware peripheral emulation by replacing BSP and device drivers, which can dynamically change the firmware to adapt to different target hardware automatically. 

Therefore, we propose a novel technique called \textit{static binary-level porting} which meets the above requirement. This method treats firmware re-hosting as porting the firmware to the boards supported by the emulator. Three observations inspired us to make this choice.

1. Emulators like QEMU support firmware re-hosting on multiple processors, and it supports full emulation of a few boards, such as \texttt{versatilepb}, \texttt{vexpress-a9}, \texttt{i440FX}, etc. These boards contain a large number of peripherals, such as UART, Ethernet, I2C, etc. The firmware is easy to emulate if we port them to these boards.

2. The modern RTOSes have a good design for easy porting. Usually, they contain a device driver infrastructure for easy driver development. This design greatly separates driver codes from the hardware. Developers only need to rewrite the BSP routines and insert new driver codes into the device driver layer when porting the firmware from one board to another. 

3. The existing emulators like QEMU can emulate various hardware and provide enough execution/memory fidelity. The response will be the same as that on real hardware. Therefore, security analysts can conduct vulnerability assessments like developing exploits/patches in a full emulated environment, and these exploits/patches can be directly applied to the firmware of a real device.


\section{Design}
\subsection{Overview}
\begin{figure*}[t]
\centering
\includegraphics[scale=0.7]{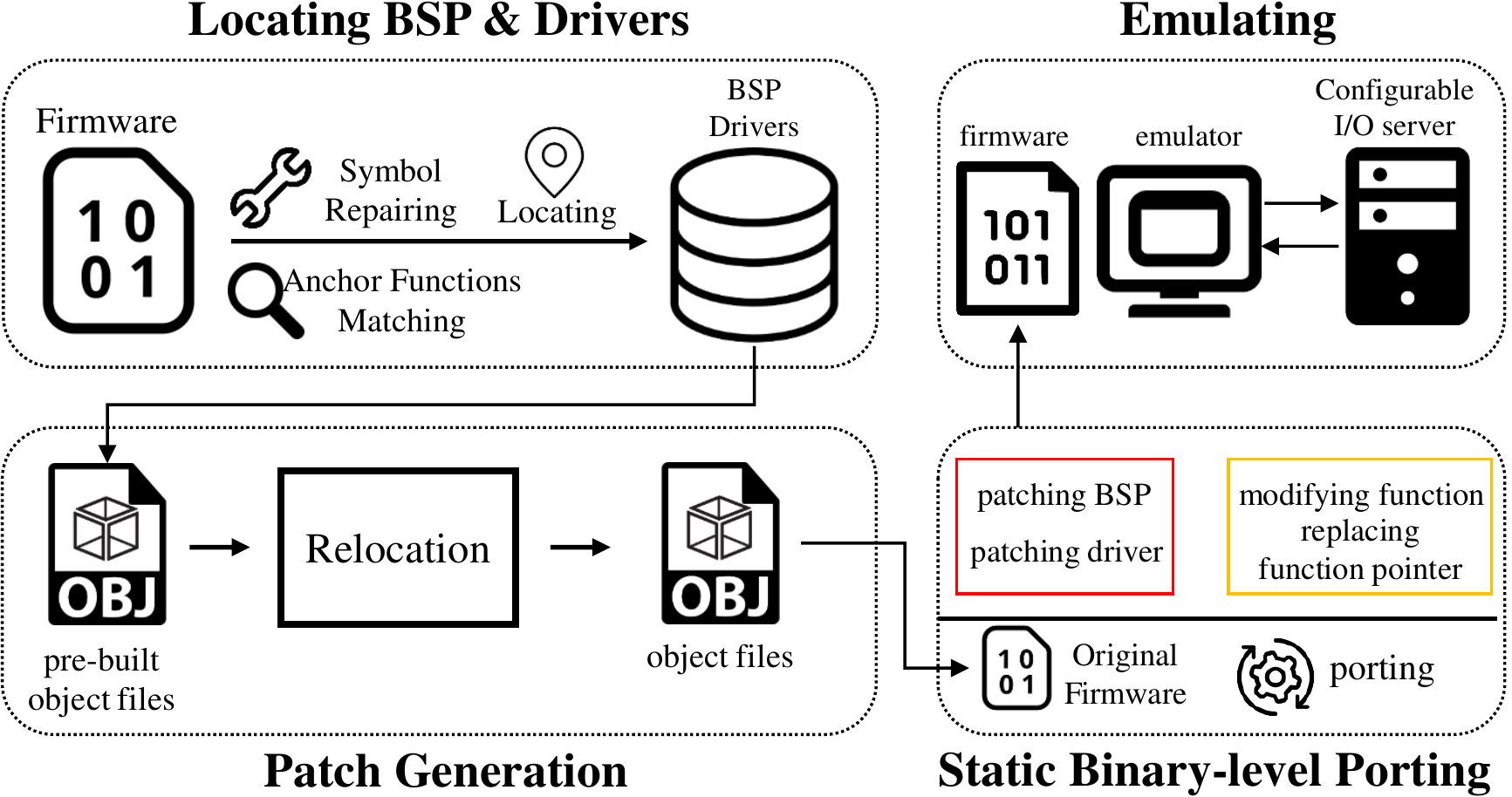}
\caption{Solution Overview}
\label{figoverview}
\end{figure*}

\noindent \textbf{Goals}

To address the above problems, our goal is to build a tool to patch the firmware thus we can make it run correctly in emulators for dynamic analysis. In particular, the aim is to re-host firmware with widely deployed RTOSes including VxWorks, RT-Thread, NuttX, and Zephyr. Specifically, we have the following goals:

1. \textit{High Adaptation. }
The tool can automatically patch the firmware and re-host them using existing emulators. 

2. \textit{High Fidelity. }
The method should have sufficient fidelity that PoCs or exploits developed under the emulated environment can be applied to the real device.

\vspace{1mm}
\noindent \textbf{Prerequisites:}

1. We have a development environment for the RTOS used in the firmware, along with a cross-compilation toolchain for the target processor. Most RTOSes are open-source and support development on multi-architecture by default.

2. We have an emulator with several boards supporting different types of peripherals. Fortunately, emulators like QEMU have plenty of boards and a few of them support multi-type peripherals like USART, I2C, Ethernet, etc.

\vspace{1mm}
\noindent\textbf{Solution Overview}

We outline the design of the proposed approach. As shown in Fig.\ref{figoverview}, the overall workflow consists of three steps. \textbf{(1) Locating BSP and Drivers.} We identify the \textit{Anchor Function} in the firmware through symbol repairing and binary similarity matching, then locate the BSP and drivers used in the firmware. \textbf{(2) Patch Generation.} We generate an object file that contains "correct" BSP routines and drivers, which means these routines and drivers can function as expected on the emulator's board. Then we relocate symbols in the object file to produce a complete patch according to the target firmware. \textbf{(3) Static Binary-level Porting.} We apply the patch to the firmware. Note that the patching strategies vary according to the target CPU architecture. Besides, we implement a \textit{Configurable I/O Server} to provide external input to the emulator.

We implemented our technique in a tool called \textit{FirmPorter}, which consists of 210 lines of Python code for locating BSP and drivers, 1046 lines of Python code for patching and integration. We also write 95 lines of IDAPython\cite{IDAPython} scripts for firmware information extraction. The patching code mainly utilizes Unicorn\cite{unicorn} for binary rewriting and LIEF\cite{LIEF} library for ELF file manipulation. 

\subsection{Locating BSP and drivers}
The prerequisite of porting RTOS at binary level includes 3 important questions: 1. Where are the BSP routines? 2. Where are the initialization functions of drivers? 3. What drivers exist in the firmware? To answer these questions, we first introduce the concept of \texttt{Anchor Function} in FirmPorter, and then we solve the problem of identifying Anchor Functions in stripped binaries. At last, we find BSP and drivers through Anchor Functions. 

\begin{figure*}[hbt]
\centering
\includegraphics[scale=0.55]{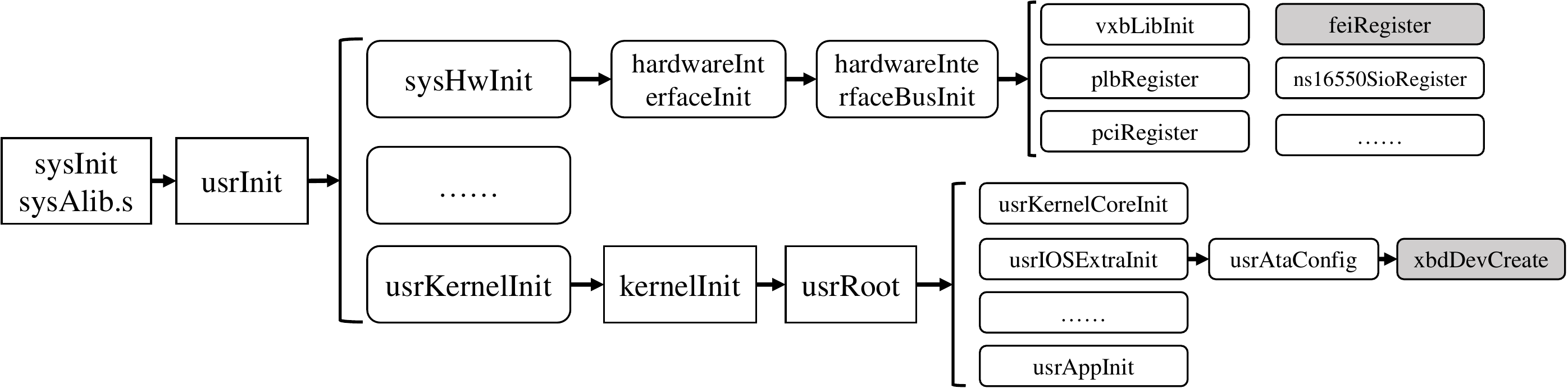}
\caption{Boot sequence of VxWorks}
\label{figvxboot}
\end{figure*}

\begin{figure*}[hbt]
\centering
\includegraphics[scale=0.55]{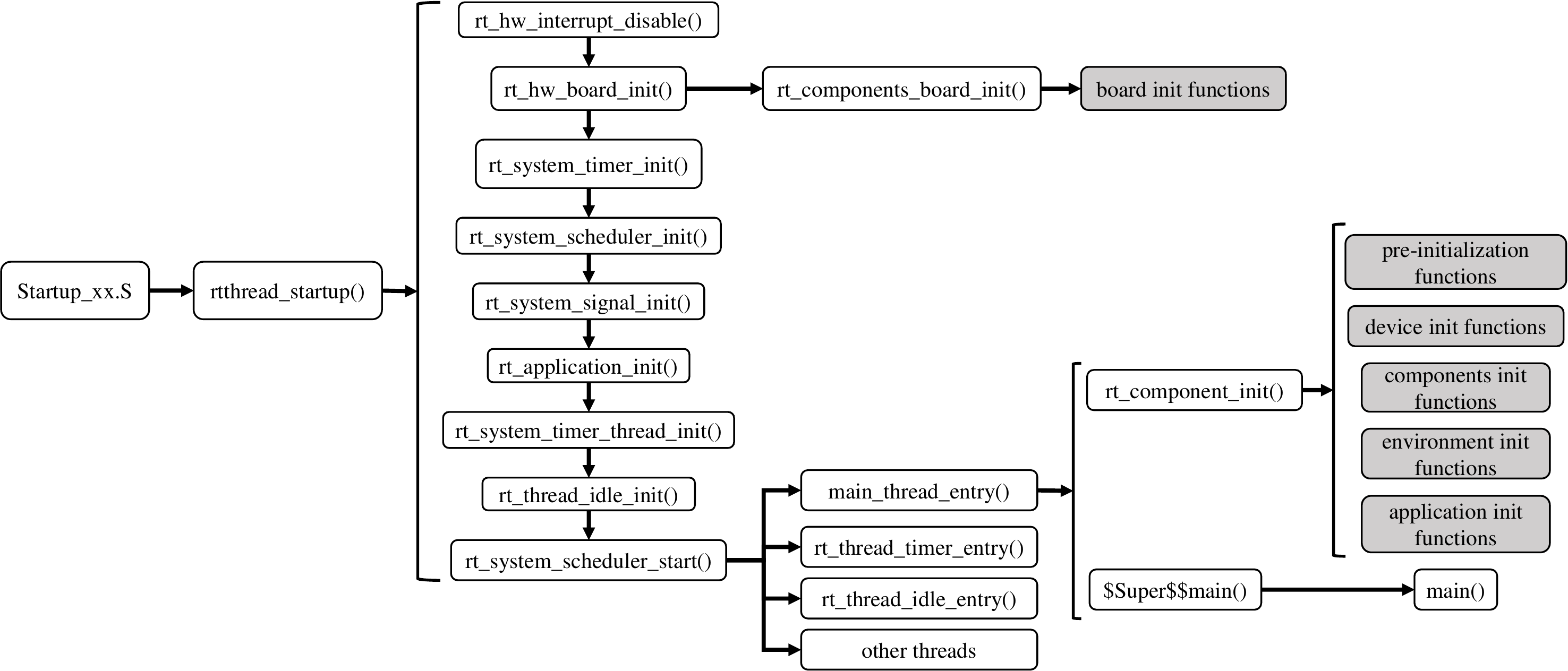}
\caption{Boot sequence of RT-Thread}
\label{figrtboot}
\end{figure*}

\vspace{1mm}
\noindent \textbf{Anchor Function. }
In a well-designed RTOS, BSP routines and drivers are initialized in specific functions. For instance, from Fig.\ref{figrtboot} we can see that device drivers are registered in \texttt{rt\_components\_board\_init} and \texttt{rt\_components\_init} in RT-Thread. Things get a little bit complicated in VxWorks. For instance, in Fig.\ref{figvxboot}, VxBus-based drivers\cite{vxdrv1} like \texttt{fei} network driver are registered in \texttt{hardwareInterfaceBusInit}, while Legacy Drives like disk driver \texttt{ata} are registered and initialized in \texttt{usrAtaConfig}. As for BSP routines, we roughly categorize the binary code executed before the kernel as BSP code, which is a superset of BSP routines. In VxWorks, we consider the code executed before \texttt{usrRoot} as BSP code.

We define the functions similar to that listed above as \textit{Anchor Function} because we can locate drivers or BSP when determining these functions. These functions exist for the abstraction and decoupling of the RTOS. Under normal circumstances, developers only modify the contents of the Anchor Functions without changing themselves.

\vspace{1mm}
\noindent \textbf{Identifying Anchor Functions in Stripped Binaries. }
We know exactly which function is Anchor Function if we have symbols of the firmware. However, two situations are often encountered in practice: 1. The header information is removed when the firmware is released, but symbols are retained in a certain part of the firmware. 2. The symbol information is removed before the firmware is released. We solve these problems individually.

\textit{a) Repairing Symbols. }
In RTOS like VxWorks, the system needs to use symbol information when it is running. The symbol table is initialized in the function \texttt{usrSysSymTblInit} and filled into the \texttt{symTbl} structure. In most cases, the symbol information is placed at the end of the firmware, which is stored in a certain format. For example, VxWorks 6.x stores the symbols of functions in the following structure:

\lstset{language=C}
\begin{lstlisting}
typedef struct symbol
{
    SL_NODE     nameHNode;
    char        *name;
    void        *value
    UINT32      symRef;
    UINT16      group;
    SYM_TYPE    type;
} SYMBOL;
\end{lstlisting}
	
We can implement a simple algorithm to recover symbols hidden in the firmware. The related IDAscript is listed in the Appendix. In this case, we can directly identify the Anchor Functions.

\textit{b) Lightweight Anchor Function Matching Algorithm. }
An Anchor Function always has a fixed caller function. Besides, its adjacent functions, which have the same caller function as the Anchor Function, have possibilities to be in a specific set of functions but not others. For example, in Fig.\ref{figvxboot}, if we can determine \texttt{usrKernelCoreInit}, then we are 100\% sure that its caller is \texttt{usrRoot}. Take \texttt{usrRoot} as a start point, it is natural to infer its callee functions have a certain probability of being \texttt{usrIOSExtraInit} or \texttt{usrAtaConfig}, and impossible to be \texttt{hardwareInterfaceInit} or other functions.

The underlying truth is the Modular Programming design in RTOS. When a developer wants to add or remove a driver in RTOS, she or he would not change the code related to the RTOS, but source code of the driver and configuration files only. For instance, in VxWorks developers add or remove a driver by enabling specific Macros, while in RT-Thread they use menuconfig to enable or disable a functionality. The driver code will be inserted into a fixed location in the RTOS, and the other part of the RTOS will remain the same.

We now define \textit{extended Anchor Functions} as functions that indispensable for RTOS, which are usually responsible for initialization. For instance, in Fig.\ref{figrtboot}, \texttt{rtthread\_startup} and all its callee functions are extended Anchor Functions. Besides, \texttt{rt\_components\_board\_init} and \texttt{main\_thread\_entry} also belong to extended Anchor Functions. These functions are fixed in source code in RTOS and developers will not change them when developing.

Based on this observation, we propose an improved \textit{Lightweight Anchor Function Matching Algorithm}. We first use Diaphora\cite{diaphora} to perform traditional binary matching on two firmware samples and select only the result of \textit{Best Match} and \textit{Partial Match} with a score greater than 0.75. Although there are a few matching results, it can be found that Diaphora has a very high accuracy for identifying extended Anchor Functions and can be considered as a \textit{Perfect Match} after many tests. Next, we use the pre-prepared RTOS CFG knowledge base to infer the caller of the function in the matching result. Then, we use the BB number, CFG, etc. to match the callee functions in the inferred caller function. During the matching process, we can set the matching confidence to be very high (e.g., 95\%). In most cases, the correct function matching score is close to 1, while the unmatched function has a low similarity score.

\begin{algorithm}[h]
  \caption{extended Anchor Function Matching}
  \begin{algorithmic}[1]
 	\State $t = FirstMatching$
 	\State $T = \varnothing$
 	
 	\While {$t$ is not empty}
 	 \State $T=T\cup t$;
 	\State $t = \varnothing$

    \ForAll {$s$ such that $s\in t$}

      \State $t=t\cup Caller(s)$;

    \EndFor
   
       \EndWhile

    \ForAll {$s$ such that $s\in T$}
   
	$//$ DFS means Depth-First-Search

    \State $DFS(s)$ and do simple matching; 
    
    \EndFor
  \end{algorithmic}
\end{algorithm}

The algorithm works well because we restrict the potential matching results into a small range, thus reducing the collision or mismatching results. Although our matching algorithm is somewhat naive where it may fail in more complicated situations, in the actual scenario, we can further add constraints (e.g., a function A registered in an unknown driver function B is called by a socket function, then the function B can be judged as a initialization function of an Ethernet adapter), or just compare the IR of the function, where the RTOS knowledge base can be continuously iterated to improve the accuracy of function matching.

\vspace{1mm}
\noindent \textbf{Locating BSP and drivers. }
The goal of determining the Anchor Function of the RTOS is to locate the BSP and drivers. Locating BSP is simple. We treat the code executed prior to the specific Anchor Function, e.g., \texttt{usrRoot}, as BSP. When dealing with drivers, we found that there are always error messages existing in the driver for debugging convenience, which will not be removed even after the driver is released. This is extremely common in commercial embedded products, which have a warranty and the error messages can help the maintenance engineer quickly locate the problems. Based on this fact, we design a lightweight strategy to locate drivers. We traverse the Call Graphs of driver initialization routines to match the certain strings in error messages. The goal of locating drivers is not to determine the specific driver like LAN91C111 Ethernet controller driver, but the type of the driver. For example, if we encounter a "socket error", we consider this driver as an Ethernet driver then finish matching.

\subsection{Patch Generation}
An important step before static binary-level porting is generating patches that contain BSP routines and drivers that can be correctly run by the emulator. In this section, we first discuss the process of generating patches from source code. Although this part requires human effort, it only needs to be done once and these patches can be applied to various firmware. Then we relocate the patches to adapt to the firmware.

\vspace{1mm}
\noindent\textbf{Patch Compilation. }
To generate patches containing correct BSP routines and drivers, we need to determine the \textit{template board}, to which we port the firmware. Our strategy is to port firmware running on the same processor to the same template board. However, if the emulator supports more than one board with the same CPU architecture, we will port the firmware to the board with the same series, such as STM32 and NXP.

The patch is a piece of binary code that modifies or rewrites problematic functions in firmware, and it will be loaded into memory at runtime. 
Usually, the base address of firmware is specified at compile-time, and the heap range is set in BSP. Therefore, it is difficult to determine which range of memory is free at runtime after firmware compilation. This also makes it hard to determine the address to load the patch to prevent memory overlapping.

Normally, firmware generation includes four stages: preprocessing, compilation, assembly, and linking. At the beginning of the linking stage, the object file contains external symbols that need to be relocated. Based on this fact, we generate an object file that contains correct BSP routines and drivers without linking, and we do the relocation works independently after determining the base address of the object file.

An interesting fact is that the GNU toolchain or its modification is widely used in most modern RTOS development. A traditional GNU project uses a Makefile to manage the compilation process and uses custom link scripts for linking. Based on this fact, we extract building relationships between source files by parsing each Makefile in RTOS's source code folders, then generate object files that contain correct BSP routines and drivers by following the building process described in Makefile. Finally, these object files will be linked into a large object file as the patch, which needs further adaptation.

\vspace{1mm}
\noindent\textbf{Patch Adaptation. }
External symbols in the patch are usually library functions like \texttt{malloc}, \texttt{bcopy}, \texttt{memcpy}, etc. To relocate these symbols, the patch's base address needs to be determined so that we can calculate the offset between the function to be patched and the patched function. In fact, we can determine the address range of \texttt{.text} segments, \texttt{.data} segments, \texttt{.bss} segments, etc. of the original firmware by parsing the firmware, which is usually an ELF file. These segments contain important data structures, such as global variables and method tables. Therefore, the patch must be put after these segments. However, heap space comes after these segments and occupies a huge memory space. Since we do not know the size of the heap, the patch is likely to be overwritten by the heap operation. In most cases, the range of the heap space is usually set in BSP. For example, in VxWorks, \texttt{sysMemTop} sets the boundary of the heap space. The patch can be put after the \texttt{sysMemTop} to ensure the integrity of the patch at runtime.

We specify the patch's base address and external functions' address in a script for linking to generate a complete patch. The patch can be loaded directly into the memory without any further modification.

\subsection{Static Binary-level Porting}
The core of static binary-level porting is redirecting the control flow from original functions to the correct ones in the patch. We first introduce two common patching strategies that are applied to binaries, then we solve the communication problems between the firmware and the host when peripherals are not supported by the emulator.

\vspace{1mm}
\noindent\textbf{Patching Strategies}

\textit{a) Direct Function Modification. }
FirmPorter modifies the first instruction of the function in the firmware to jump to a new one in the patch. For example, a driver initialization function named \texttt{usrAtaConfig} in firmware cannot be emulated correctly, and we have a new \texttt{usrAtaConfig} in the patch, which can be emulated without errors. Therefore, we change the first instruction of \texttt{usrAtaConfig} from \texttt{push ebp} to \texttt{jmp 0xc000120}, where \texttt{0xc000120} is the address of \texttt{usrAtaConfig} in the patch.

When dealing with x86-based firmware, \texttt{jmp} instruction is used to jump to an absolute address. In particular, the \textit{far jump} instruction is used to go to any location within 0x00000000$\sim$0xffffffff, where the machine code of \textit{far jump} is \texttt{0xea}.

When dealing with Arm-based firmware, we use different strategies according to the feature of instructions. Take \textit{AArch32} as an example, there are four unconditional jump instructions in Arm: \texttt{B}, \texttt{BL}, and the corresponding \texttt{BX}, \texttt{BLX}, where the last two change the working mode between \textit{Arm} and \textit{Thumb}. \texttt{B} is followed by an offset, which is equivalent to \texttt{MOV PC, [PC+offset]}, similar to the \texttt{jmp} instruction in x86. \texttt{BL} is also followed by an offset, which is equivalent to \texttt{MOV LR, BP MOV PC, [PC+offset]}, similar to the \texttt{call} instruction in x86. According to the chip's architecture, the working modes of Arm include \textit{Arm mode} and \textit{Thumb/Thumb-2 mode}. In \textit{Arm mode} the address space is $\pm$32M while in \textit{Thumb mode} the address space is $\pm$4M. Therefore, if heap space is too large, the code in the patch will far exceed the \texttt{B} and \texttt{BL}'s address space, which means that we cannot simply jump from one function to another. 

\begin{figure}[t]
\centering
\includegraphics[scale=0.5]{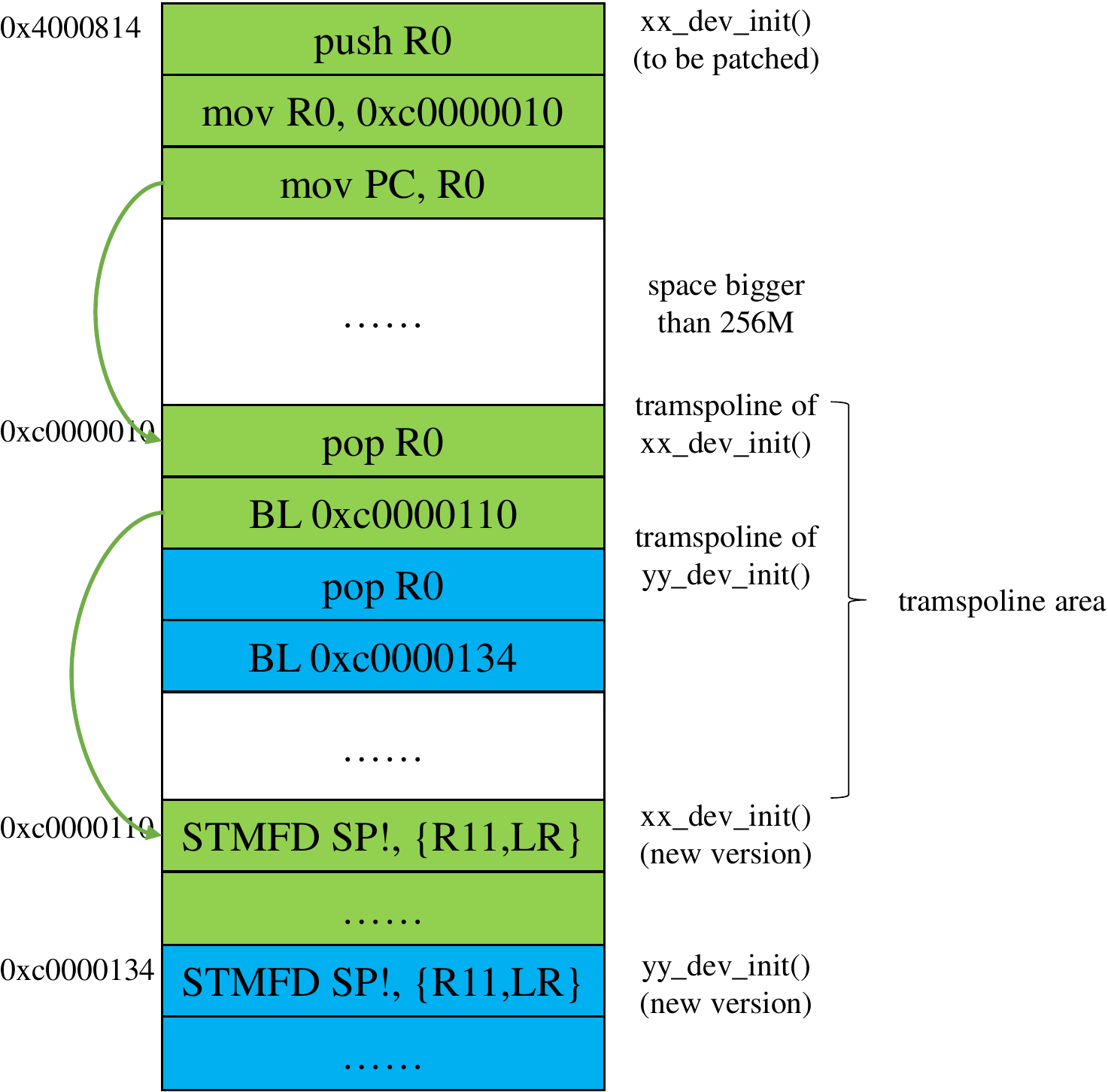}
\caption{Static binary-level porting on Arm processor with trampolines}
\label{figjumpsled}
\end{figure}

We use an indirect jump scheme to solve this problem, which uses a general-purpose register (such as \texttt{R0}) as a stepping stone. As illustrated in Fig.\ref{figjumpsled}, FirmPorter saves the value of the register \texttt{R0}, then uses \texttt{MOV} instruction to copy the value of \texttt{R0} to the register \texttt{PC} so that we can jump to an absolute address. Then it creates a \textit{trampoline area} before the patch, where each part contains a trampoline. In a trampoline, FirmPorter restores the value of \texttt{R0} and calls the correct function.

The next step is to concatenate the patch and the firmware. FirmPorter generates the patch in binary format by calling \texttt{objcopy} command in the toolchain. Then it fills the bss segment, heap space, and other non-allocated areas with zero. Afterward, FirmPorter concatenates the patch and the firmware by simply calling \textit{cat} command. Note that FirmPorter needs to allocate an area for \textit{trampoline} when dealing with firmware running on an Arm processor.

\textit{b) Function Pointer Replacement. }
To achieve a high-level abstraction, some RTOSes use pointers to call the initialization function. For example, in RT-Thread, the \texttt{rt\_component\_init()} function will traverse the \texttt{\_\_rt\_init\_rti\_end} area and call the initialization function through pointers in turn. We can achieve the purpose of replacing drivers by directly modifying the pointer in this area to point to the patch function at the end of the firmware. Of course, we can also modify the function to hijack the control flow from the original driver initialization function to the patched one, but directly modifying the pointer has two advantages: 1. It reduces the modification of the original firmware without unnecessary trampolines. 2. During analysis, the control flow graph will be smoother and easy to analyze.

\vspace{1mm}
\noindent\textbf{Driver Hacking \& Configurable I/O Server. }
Our goal is to establish a vulnerability verification environment that resembles the real device, so an input/output similar to the physical environment is required. Since firmware contains a large number of peripherals and the communication approaches between the emulator (such as QEMU) and the outside world are limited (Ethernet, serial port, I2C, etc.), it is important to systematically solve the problem of data communication inside and outside the emulator. FirmPorter does not modify the emulator, it can only use the data communication approaches already supported by the emulator. Based on this fact, we re-designed a set of communication strategies.

Our solution is to install an equivalent driver into the RTOS, which employs the functionalities supported by the emulator only. This solution is suitable when the vulnerability does not occur in the driver and the analyst just wants to pass the payload to the destination with high integrity.

We propose a \textit{Driver Hacking} method to generate drivers with equivalent data transmission capability. We use transmitting/receiving methods supported by the emulator to re-implement the control-related functions and the communication-related functions. For example, we can utilize \textit{semihosting}\cite{semihosting} technique supported by qemu-system-arm when dealing with Arm-based firmware. \textit{Semihosting} supports the virtual machine to read/write the host's stdin, stdout, stderr, and files. We employ functions of reading/writing files to exchange data between emulators and hosts, and stdin/stdout channels to transmit flags. The flags are used to indicate when to start or stop data transmission. When dealing with x86-based firmware that supports serial controllers or Ethernet controllers, we can reuse the Serial channel or Ethernet channel to exchange data and control transmission processes.

External input is an important issue. When re-hosting firmware for fuzzing, all inputs are random after mutation. In this work, we re-host firmware for vulnerability assessment, and there are two types of external input: 1. Data that needs to be tested, such as abnormal input, exploits, etc.2. Data that supports the firmware performing normally. For example, a PLC firmware whose purpose is to monitor the boiler pressure must read the boiler pressure at regular intervals to determine whether it exceeds a certain threshold. When the pressure exceeds the limit, the PLC performs a series of pressure reduction operations. Therefore, if the value is not properly provided during the firmware emulation, the performance of the emulated firmware will not resemble the actual one.

An embedded device with a specific RTOS is usually applied in a specific environment, which behaves differently from that with a General Purpose Operating System (GPOS). Therefore, we implement a \textit{Configurable I/O Server} to provide different scenarios that similar to the real world. We have written a series of configuration scripts for different scenarios. In this case, when the I/O server is started, it will interact with the PLC and provide the value of pressure through the ADC interface regularly. 

It should be noted that the purpose of several emulation works is to trigger more execution paths when re-hosting firmware, so the output of the firmware is actively ignored (the output result does not affect the execution path). However, our goal is to build a firmware emulation analysis environment that resembles the real environment as much as possible. Besides, an abnormal output also indicates that the firmware is under attack, so we still deal with the output of the firmware.

\section{Evaluation}
We evaluated FirmPorter from three different angles: (1) whether it can successfully identify the Anchor Functions and locate the BSP and drivers used in the firmware; (2) whether it can successfully emulate firmware; (3) whether it provides enough re-hosting fidelity that PoCs/exploits developed by emulation can be applied on real devices.

Our targets include 11 firmware with 4 types of RTOS (VxWorks, RT-Thread, Nuttx, and Zephyr) and run on 5 different processors (x86, Arm Cortex-M3/M4, Arm1176, Arm926EJ-S, and MIPS32). We re-host firmware via 3 open-source emulators: QEMU, xpack QEMU Arm, and QEMU for MIPS. The RTU and PLC samples are based on VxWorks, while the wireless temperature sensor network samples are based on RT-Thread. These firmware run on AMD board, SPEAr 680 board and STM32 board, respectively. We also build 1 firmware runs on the STM32 board, 2 firmware run on the NXP board, 1 firmware runs on the Samsung board, 1 firmware runs on the RaspberryPi board, and 1 firmware runs on the PIC32 board. These firmware contain different vulnerabilities. Table \ref{tbtargets} lists detailed information of the targets.

\subsection{Results of Locating Anchor Functions and Drivers}

We first explore the effectiveness of the algorithm to locate the Anchor Functions and drivers used in the firmware. Among the firmware we tested, there are 6 firmware that contain a complete symbol table. We first preprocessed these firmware samples and removed the symbol tables. For NAS firmware, we found that the symbolic information of the firmware could not be removed using the traditional \texttt{strip} tool, so we chose to directly destroy the ELF header of the firmware. This is also a common situation encountered in actual scenarios.

\vspace{1mm}
\noindent\textbf{Locating Anchor Functions. }
We take the existing firmware with the same CPU architecture as the "first program", and the stripped firmware sample as the "second program", using the Diophora for binary comparison. Then we use the context matching algorithm combined with the matching result to perform a secondary identification of the Anchor Function. For NAS firmware, we directly use the symbol repairing algorithm to restore the symbol table. The results are listed in Table \ref{recoveryresults}.

\begin{table}[htbp]\scriptsize
\centering
\begin{threeparttable}[b]
\caption{Results of Locating Anchor Functions}
\begin{tabular}{p{2cm}<{\centering}p{0.7cm}<{\centering}p{0.5cm}<{\centering}p{0.3cm}<{\centering}p{0.7cm}<{\centering}p{0.7cm}<{\centering}}
    \toprule
        Firmware		&Syms &DM &AF &M &AFAI\\
    \midrule\midrule
    		Music Player	  &532 &120 &1 &4 &5\\
    		Car	 &1031 &226 &1 &4 &5\\
    		Web Server	 &811 &82 &0 &5 &5\\
    		UART Server	  &1075 &413 &1 &4 &5\\
    		TCP Echo Server	  &732 &75 &0 &4 &4\\

    \bottomrule
\end{tabular}
\label{recoveryresults}
 \begin{tablenotes}
        \footnotesize
        \item[*] \textbf{DM} means Direct Matching, \textbf{AF} means Anchor Function, \textbf{M} means Missing, \textbf{AFAI} means Anchor Functions After Iteration
        \
      \end{tablenotes}
      \end{threeparttable}
\end{table}

\textit{Direct Matching} is the result of matching by Diophora. It can be seen that the number of functions in the first matching is relatively small. This is because we need to ensure that the result of the first matching is 100\% correct. Only the results of \textit{Best Match} and \textit{Partial Match} with scores greater than 0.75 are counted. In fact, if we include all the results in the \textit{Partial Match} and the results in the \textit{Unreliable Match} (a few matching results are correct), the number of matching functions for the first time increases by about 20-40\%. After manual checks, we confirm the result of the first matching is completely correct. It can be seen from the matched functions that some Anchor Functions are not identified. For example, an Anchor Function \texttt{rt\_component\_board\_init} is not matched in this round. This function is responsible for traversing the driver function pointer for initialization. By observing the function structure, it can be found that the core structure of the \texttt{rt\_component\_board\_init} function is a for loop, which traverses and calls the function according to the address range. However, there are too many similar functions with the same CFG structure, which are difficult to identify. And unfortunately, the caller of \texttt{rt\_hw\_board\_init} cannot be recognized either. However, we correctly identify the \texttt{rt\_system\_heap\_init} function. According to the Call Graph generated previously, we can determine that the caller of this function is \texttt{rt\_hw\_board\_init}. Therefore, the potential function set of this function can be determined. Under this condition, we can easily identify the function based on the characteristics of the \texttt{rt\_component\_board\_init} function (CFG, BB, etc.). From the table, we can see that in the second matching round the result contains all the  Anchor Functions of RTOS. In addition, we repairs the symbol table of the NAS firmware, and the final repairing result is the same as the actual symbol table as expected.

\begin{table}[htbp]\scriptsize
\centering
\begin{threeparttable}[b]
\caption{Results of Symbol Repairing in the firmware with VxWorks}
\begin{tabular}{p{1cm}<{\centering}p{1cm}<{\centering}p{1cm}<{\centering}p{1cm}<{\centering}p{0.7cm}<{\centering}p{0.7cm}<{\raggedright}}
    \toprule
        Firmware		&Syms &Repaired &Rate &TP \\
    \midrule\midrule
    		NAS	  &7798 &6630 &85\% &100\%\\

    \bottomrule
\end{tabular}
\label{recoveryresults}

      \end{threeparttable}
\end{table}

We use this method to process 5 real-world firmware. The first problem is that the ELF header of the firmware is damaged, or the firmware is a pure binary file. Therefore, we must first be able to load the firmware correctly and generate a CFG diagram. We use the method in \cite{basnight2013firmware} to identify the function information in IDA according to the function prologue and epilogue to generate a CFG. For the two VxWorks firmware, we have repaired most of the function symbols, and the Anchor Functions identified are correct. For the remaining 3 WSN firmware, we can see from the recovery results that the Anchor Functions are also correctly identified.

\begin{table}[htbp]\scriptsize
\centering
\begin{threeparttable}[b]
\caption{Locating Drivers in Stripped Binaries}
\begin{tabular}{p{2cm}<{\centering}p{1.5cm}<{\centering}p{0.9cm}<{\centering}p{0.9cm}<{\centering}}
    \toprule
        Firmware		&RTOS	&Drivers &Identified\\
    \midrule\midrule
    		Music Player	 &RT-Thread  &2 &2\\
    		NAS	 &VxWorks  &3 &3\\
    		Car	&RT-Thread  &3 &3\\
    		Web Server	&Nuttx  &4 &4\\
    		UART Server	&Zephyr  &1 &1\\
    		TCP Echo Server	 &Nuttx  &2 &2\\

    \bottomrule
\end{tabular}
\label{recoveryresults}

      \end{threeparttable}
\end{table}

\begin{table}[htbp]\scriptsize
\centering
\begin{threeparttable}[b]
\caption{Locating Drivers in Real Firmware Samples}
\begin{tabular}{p{2.0cm}<{\centering}p{1.4cm}<{\centering}p{0.5cm}<{\centering}p{0.5cm}<{\centering}p{0.5cm}<{\centering}p{0.5cm}<{\centering}}
    \toprule
        Firmware 	&RTOS	&Syms &SR &AF &Drivers\\
    \midrule\midrule
    		RTU	&VxWorks &19221 &17683 &7 &5 \\
    		PLC	 &VxWorks &15757 &13968 &7 &7\\
    		WSN-sample	&RT-Thread &405 &130 &5 &3\\
    		WSN-store	&RT-Thread &860 &212 &5 &3\\
    		WSN-upload	&RT-Thread &514 &157 &5 &3\\

    \bottomrule
\end{tabular}
\label{realfirmware}
\begin{tablenotes}
        \footnotesize
        \item[*] \textbf{SR} means Symbol Recovered, \textbf{AF} means Anchor Functions
        \
      \end{tablenotes}
      \end{threeparttable}
\end{table}

We can see from the results that VxWorks has more Anchor Functions than other RTOS. This is because VxWorks supports both Legacy device drivers and VxBus device drivers for compatibility requirements, and these drivers are registered in separate locations. For example, the disk driver initialization routine \texttt{ataXbdDevCreate()} does not initialize in the \texttt{hardwareInterfaceBusInit}, because the disk driver does not support VxBus method in VxWorks according to the manual. Therefore, the disk driver porting in VxWorks is to replacing the \texttt{ataXbdDevCreate()} with the correct one.

\vspace{1mm}
\noindent\textbf{Locating Drivers. }
We determine the Anchor Function that contains the location of the driver initialization function by traversing the Call Graphs and checking the error message strings present in that driver function, then we determine the category to which the driver belongs by comparing it with the string in our database. We classify the driver types into the following 13 categories:

\textit{GPIO, UART, I2C, SPI, ADC, PWN, RTC, WATCHDOG, Ethernet, HWTIMER, Sensor, Storage, Other}

There are many approaches to determine which peripheral drivers are included in the firmware. For example, in RTU and PLC and NAS firmware, we can not only look for common names like "fei0", "sm", "lnc0", etc. that clearly represent a NIC driver, but also by searching the firmware for strings like "inet on ethernet" or "socket is not connected" to match the "inet" and "socket", which indicates the firmware using an Ethernet adapter. We collected a large number of driver names from the open-source code and stored them in the database for the heuristic search.

Besides, we deal with a special case when locating drivers in the PLC sample, which is based on VxWorks. According to the manual, the ADC and I2C drivers are classified as "Other Classes" in VxWorks and initialized in the \texttt{usrIosCoreInit()}. They do not follow any device driver models and exist independently.

\subsection{Results of Firmware Re-hosting}
\begin{table}[t]\scriptsize
\centering
\begin{threeparttable}[b]
\caption{Firmware Re-hosting Setup}
\begin{tabular}{p{2.3cm}<{\centering}p{2.65cm}<{\centering}p{2cm}<{\centering}}
    \toprule
        Firmware	&Emulator	&Template Board	\\
    \midrule\midrule
    		RTU	&QEMU	&i440FX\\
			WSN-sample	&xPack QEMU Arm	&STM32F429	\\
			WSN-store	&xPack QEMU Arm	&STM32F429	\\
			WSN-upload	&xPack QEMU Arm	&STM32F429	\\
			Music Player	&xPack QEMU Arm	&STM32F429 \\
			PLC	&QEMU	&versatilepb \\
			NAS	&QEMU	&versatilepb	\\
			Car	&QEMU	&versatilepb	\\
			Web Server	&QEMU	&versatilepb \\
			UART Server	&xPack QEMU Arm	&STM32F429	\\
			TCP Echo Server	&QEMU for MIPS	&pic32mz-meb2	\\
    \bottomrule
\end{tabular}
\label{tbsetup}
      \end{threeparttable}
\end{table}

In this experiment, first, we demonstrate that firmware ported by FirmPorter can run correctly in unmodified emulators. Then we explore the scalability and benefits of static binary-level porting.

\vspace{1mm}
\noindent\textbf{Re-hosting Capability. }
To evaluate the effectiveness of FirmPorter, we compare the behavior on the emulator with that on the real device. Specifically, we consider the firmware re-hosting is successful if each peripheral functionality can perform equivalently on the emulated system as on the real hardware. As firmware's functionality differs, the "success" standards for the execution of each firmware are set accordingly. First, we list the set-up of the firmware we are going to port, which is listed in Table \ref{tbsetup}.

\begin{table}[t]\scriptsize
\centering
\begin{threeparttable}[b]
\caption{Result of Firmware Re-hosting}
\begin{tabular}{p{2cm}<{\centering}p{0.3cm}<{\centering}p{0.3cm}<{\centering}p{0.3cm}<{\centering}p{0.3cm}<{\centering}p{0.3cm}<{\centering}p{0.4cm}<{\centering}p{0.4cm}<{\centering}p{0.3cm}<{\centering}}
    \toprule
        Firmware	&Ether.	&Disk		&UART &ADC &RF &GPIO &Zigbee &SD\\
    \midrule\midrule
    		RTU				&e	&d	&u	&a	&-	&g &- &s\\
			WSN-sample		&-	&-	&u	&a	&r	&- &- &-\\
			WSN-store		&-	&-	&u	&-	&r	&- &- &s\\
			WSN-upload		&e*	&-	&u	&-	&r	&- &- &-\\
			Music Player	&-	&-	&u	&-	&-	&g &- &-\\
			PLC				&e	&d	&u	&a	&-	&g &- &s\\
			NAS				&e	&d	&u	&-	&-	&- &- &-\\
			Car				&-	&-	&u	&-	&-	&g &z &-\\
			Web Server		&e	&-	&u	&-	&-	&- &- &s\\
			UART Server		&-	&-	&u	&-	&-	&- &- &-\\
			TCP Echo Server	&e	&-	&u	&-	&-	&- &- &-\\
    \bottomrule
\end{tabular}
\label{tbrresults}
\begin{tablenotes}
        \footnotesize
        \item[*] \textbf{e} means one can communicate the emulator via TCP/IP, \textbf{e*} means one can communicate the emulator via the I/O Server, \textbf{d} means firmware can r/w files on the host, \textbf{u} means one can communicate with firmware through UART on QEMU, \textbf{a}, \textbf{r} and \textbf{z} means firmware can communicate with the I/O Server in specific protocols, \textbf{g} means firmware can r/w binary data from/to the I/O server, \textbf{s} means firmware can r/w data from/to the SD Card emulated by the QEMU.
        \
      \end{tablenotes}
      \end{threeparttable}
\end{table}

\textit{Target Board} is the template board to which the firmware is ported. We can see that all STM32-based firmware samples are ported to \texttt{STM32F429} development board, while all Arm7-based firmware samples are ported to \texttt{versatilepb} board.

The correct emulation of peripheral functions depends on the emulator we use and the development board we port. In our experiments, UART are supported on every emulator and every development board, while Ethernet and GPIO are supported on a few development boards, and RF and Zigbee are not supported. We can use the Driver Hacking method to reuse the existing peripheral functions for communication, which also means we no longer study the security issues among the replaced drivers. In our experiments, ADC, RF, and Zigbee are implemented by reusing the UART, and the GPIO of RTU, PLC, and Car are also implemented by reusing the UART. Since xPack QEMU Arm supports few peripheral functions, when re-hosting the WSN-upload sample, we use Semihosting mode to transfer data via files to provide Ethernet functionality.

To prove that FirmPorter can adapt the firmware to the hardware correctly, we test each peripheral functionality respectively. For example, to test the Web Server sample, we put different web pages to the fake SD Card and visit every HTML file from the browser correctly. Besides, we log onto the \texttt{nsh} shell from the serial terminal and find we can list the running tasks and execute other commands successfully. Therefore, we consider the "peripherals" to behave correctly after porting. The re-hosting result and the standard for success are listed in Table \ref{tbrresults}.

An interesting case is the Car sample. As Zigbee and PWM are not a general peripheral class in our category, it is automatically replaced with a reused UART driver. To testify the peripheral functions, we analyze the commanding protocol and implement it in the script, which will be used by the Configurable I/O Server. We specifically examine the "PWM data"(sequence of 0s and 1s which are generated by the driver via Driver Hacking) and find the data stream is as expected.

The RTU is the most complicated firmware among the test cases, which has 19221 functions. It provides hundreds of protocol conversion functionalities, and we cannot test all of them. As the primary goal is to test URGENT/11 vulnerabilities on the emulated firmware, we define "success" as "correct execution from bootloader to the start of application". In the experiment, we can log into the system through the default telnet service and visit the web pages located on the virtual disk. Moreover, the RTU application starts correctly and the ADC driver behaves as expected in the application, which uses the Driver Hacking method.

\vspace{1mm}
\noindent\textbf{Scalability and Human Effort. }FirmPorter is able to port firmware with similar CPU architectures to a specific architecture, using the same BSP routines and drivers. In our experiments, all 5 Arm Cortex-M3/M4 firmware samples were ported to the STM32F429 Development Board, which belongs to the Arm Cortex-M4 architecture. In fact, the differences between Arm Cortex-M3/M4 little, e.g. Cortex-M4 has more features such as floating point calculations than Cortex M3, and the memory layout is slightly different. Since the RTOS follows a standardized and structured development approach, the access to specific registers in the application layer is encapsulated within driver functions, so the driver replacement will not have any impact on the application layer. In addition, the unified BSP replacement allows the firmware to be adapted to the STM32F429 Development Board. This is the same for RT-Thread, Nuttx, and Zephyr. In addition, the VxWorks firmware requires a bootloader, and we can boot any VxWorks-based firmware running on this development board with an all-in-one bootloader only.

In fact, most of the manual work was focused on developing RTOS-specific BSP routines, drivers and bootloaders for a specific board. In our experiments, we collect official BSP projects, open-source projects on GitHub and adapt 13 types of peripheral drivers to different RTOS. In this process, debugging took up much time. However, when this work was done, the re-hosting process for the firmware can run automatically and efficiently.

\subsection{Vulnerability Verification}
\begin{table}[t]\scriptsize
\centering
\begin{threeparttable}[b]
\caption{Exploitation Reproducibility}
\begin{tabular}{p{2cm}<{\centering}p{1cm}<{\centering}p{3.3cm}<{\centering}p{0.3cm}}
    \toprule
        Firmware	&Vuln	&Vuln func		&R\\
    \midrule\midrule
    		RTU	&Urgent/11	&iptcp\_usr\_get\_from\_recv\_queue()	&Y\\
			WSN-sample	&BOF	&rd\_tmp\_entry()	&Y\\
			WSN-store	&BOF	&nrf24l01\_recv()	&Y\\
			WSN-upload	&BOF	&net\_upload\_data()	&Y\\
			Music Player	&BOF	&header\_parse() &Y\\
			PLC	&RCE	&*** &Y	\\
			NAS	&Urgent/11	&iptcp\_usr\_get\_from\_recv\_queue()	&Y\\
			Car	&BOF	&radio\_recv()	&Y\\
			Web Server	&BOF	&simple\_xml\_parse() &Y\\
			UART Server	&BOF	&data\_read()	&Y	\\
			TCP Echo Server	&BOF	&data\_read()	&Y\\
    \bottomrule
\end{tabular}
\label{tbresults}
\begin{tablenotes}
        \footnotesize
        \item[*] \textbf{R} means Reproducibility
      \end{tablenotes}
      \end{threeparttable}
\end{table}

In this experiment, we demonstrate that FirmPorter provides enough re-hosting fidelity that PoCs/exploits developed by emulation can be applied on real devices. Then we explore whether the fidelity is enough when verifying vulnerabilities using the firmware ported by the FirmPorter.

\subsubsection{Vulnerability Exploitation. }
\textit{Vuln} shows that the vulnerabilities (Buffer Overflow, BOF) are built into each firmware except for the real-world ones. We do not set the NX which enables attackers to execute codes on stack or heap. The goal of exploiting is to output the string "exploited!" through UART by exploiting the vulnerability. For the two VxWorks-based systems, we test the CVE-2019-12255 in URGENT/11, which is a heap overflow vulnerability.

\textit{Reproducibility} shows that whether the PoCs/exploits work on the real device. After successfully triggering each vulnerability, we send exploits to real hardware through peripherals like Zigbee instead of files. As shown in Table \ref{tbresults}, the result shows that the PoCs/exploits also trigger the vulnerabilities on the devices successfully. Besides, we send the PoCs to the two VxWorks-based firmware, and the telnet service can be corrupted successfully.

We emulated an actual firmware: Modicon M241 PLC, and verified the Stuxnet-like attack proposed by AirBus\cite{airbus}. Modicon M241 PLC uses the Codesys 3.5 platform to provide PLC service, which has been reported to have multiple CVE vulnerabilities. On the Codesys platform, the compiled PLC control logic exists in the form of Arm native code. Therefore, if the control logic is inserted with malicious binary instructions, the attacker can accomplish any potential attack.

In the research of AirBus, researchers determine the PLC memory layout through crashes and write the shellcode based on these addresses. In our experiment, the symbol table of the firmware has been restored already, and the information of CPU architecture (SPEAr 680) of the PLC is obtained through disassembling the device.Therefore, we can directly write the shellcode according to the datasheet.

The attack is to invoke the socket system call and connect to a specified IP/port to send the "exploited!" messages. We insert this part of the code into a real PLC ladder logic, then modify it to make it a legal program and download it to the firmware. The development work of the shellcode is completed in the emulation environment.

We download the program to the real PLC, and it achieves the same effect as in the emulation environment. The target PC receives the "exploited!" message.

\subsubsection{Fidelity Assessment. }
The purpose of FirmPorter is to establish an environment for dynamic firmware analysis, and we specifically care about whether the vulnerabilities are triggered as the same as on the real device. Therefore, we evaluate the fidelity from two aspects: 1. Execution path similarity in vulnerable functions 2. Memory region similarity after triggering the vulnerabilities. For data collection convenience, we testify the 5 firmware running on the development board.

\begin{table}[htbp]\scriptsize
\centering
\begin{threeparttable}[b]
\caption{Local Memory Fidelity in Vulnerable Function}
\begin{tabular}{p{2cm}<{\centering}p{1.7cm}<{\centering}p{1.2cm}<{\centering}p{1.2cm}<{\centering}<{\raggedright}}
    \toprule
        Firmware 	&Vuln Func	&SHE &SHD\\
    \midrule\midrule
    		Music Player	&header\_parse() &ece4288a  &ece4288a \\
    		Car	&radio\_recv() &1a35a8e5 &1a35a8e5 \\
    		Web Server	&simple\_xml\_parse() &e7fafbc1 &e7fafbc1 \\
    		UART Server	&data\_read() &acca0cd3  &acca0cd3 \\
    		TCP Echo Server	&data\_read() &213f4a53 &213f4a53 \\

    \bottomrule
\end{tabular}
\label{fidelity}
\begin{tablenotes}
        \footnotesize
        \item[*] \textbf{SHE} means Stack Hash in Emulator, \textbf{SHD} means Stack Hash in Device
        \
      \end{tablenotes}
      \end{threeparttable}
\end{table}

\vspace{1mm}
\noindent\textbf{Execution and memory fidelity. }
We pay more attention to the part that requires vulnerability analysis, that is, whether the code that is separated from the hardware abstraction layer, such as the protocol stack, application, or kernel, behaves the same as that running on the real device. In this case, we can develop exploits or patches on the re-hosting platform, which can be directly applied to the real device.

We propose an \textit{on-demand fidelity calculation} scheme. By specifying the functions to be analyzed, the fidelity is compared in case of triggering vulnerabilities/attacks. It is difficult to collect execution data from real devices (such as a PLC) through JTAG. To testify our solutions, we collect execution trace using the ETM module (Embedded Trace Macrocell) module on the STM32F4 Discovery development board similar to Laelap, while using "-d exec, nochain" command to get the execution path on the QEMU side. In addition, we set a breakpoint after the vulnerability is triggered. When the \texttt{PC} pointer is executed there, the register data and execution stack information will be dumped.

The samples we tested are an RT-Thread firmware with Stack-overflow vulnerabilities. In order to avoid the execution uncertainty caused by hardware interrupts and context switching, we manually disable the interrupt before the function starts and enable the interrupt after the execution is over. The recording of the execution starts after the interrupt is disabled, and the entire process of exploiting the vulnerability is recorded. We extracted contents in the task stack after the exploitation was over.

\textit{Execution fidelity. }The execution paths in the device and the emulator are completely overlapped.

\textit{Memory fidelity. }We calculate the SHA-1 of contents in the task stack after the attack and find them are the same (The value listed in Table \ref{fidelity} remains the last 8 digits only). During the test, we find some of the starting addresses of the stack are different. This is because we introduced a new driver that uses a different stack size from the original firmware. However, it is meaningless to ensure that starting address of the stack is the same, especially when it comes to verification and analysis of heap vulnerabilities. On the other hand, a mature exploit should rely as little as possible on absolute addresses (except for \texttt{.text} segment). Therefore, we use this example to prove that the execution is "similar" to that on the real device. We also compared the execution path of the remaining firmware samples, and the results are listed in Table \ref{fidelity}.

\section{Limitations and Discussion}
\label{discussion}
\subsection{Fuzzing}
The purpose of this work is to reduce the workload problem caused by modifying the emulator and to support researchers to quickly emulate the firmware and set up a workbench for dynamic analysis. However, FirmPorter does not implement the fuzzing module to discover new vulnerabilities. This is because the fuzzer based on coverage feedback guidance like AFL will count one path as two or more due to the uncertainty of the interrupts. There are two solutions to this problem: 1. FirmPorter only emulates and fuzz functions of interest to avoid the interrupts caused by the whole system emulation. 2. We modify the emulator so that FirmPorter could control the interrupts during fuzzing. For example, P$^2$IM is set to fire an interrupt every 1000 Basic Blocks, which ensures the determinism of the interrupt and the repeatability of the crash. In the following work, we will use a partial simulation scheme to fuzz the firmware.

\subsection{RTOS without Device Driver Infrastructure}
As shown in Table 1, 50\% of RTOS we investigate do not have a device driver infrastructure and follow a process-oriented design. In fact, many of them have been widely used for a long time such as FreeRTOS. The difference between the two kinds of drivers lies in whether the driver method is called by the driver instance or directly by other functions. In the former case, FirmPorter simply replaces the driver by initializing a new instance, while in the latter case it needs to replace all functions related to the driver. The challenge remains in the full identification of functions belonging to one driver.

\subsection{Firmware without Symbols - Binary Similarity Matching}
Many works has been done on binary code similarity comparing. According to Haq et al.'s work \cite{Haq2019}, the binary code similarity evolves from EXEDIFF\cite{Baker1999} to ASM2VEC\cite{Ding2019}. Nowadays binary code similarity matching has become a feature learning problem that most researchers use machine learning and neural networks to match the desired code. For example, Yu et al.\cite{Yu2020} use BERT to pre-train the binary code on one token-level task, one block-level task, and two graph-level tasks, then adopt a convolutional neural network (CNN) on CFG's adjacency matrices to extract the order information. Their results outperform state-of-art results.

\section{Related Work}
\subsection{Hardware-in-the-loop Emulation}
Several works address the problems of dynamically analyzing firmware in a hardware-in-the-loop way. They forward peripheral operations to a real device when executing hardware-related code. Avatar\cite{Zaddach2014} proposed a framework that forwards MMIO operations to a device through the JTAG channel, and it integrates it with S$^2$E\cite{Chipounov2011} to conduct concolic execution on firmware. The following work Avatar2\cite{Muench2018} optimizes the original framework and integrates with PANDA\cite{Dolan-Gavitt2015} so that it can replay recorded I/O operations without hardware. SURROGATES\cite{Koscher2015} enables near real-time dynamic analyses of firmware by providing a low-latency FPGA bridge between the host's PCI Express bus and the system. PROSPECT\cite{Kammerstetter2014} forwards peripheral hardware accessed from the original host system into a virtual machine. \cite{Kammerstetter2016} optimize the performance of forwarding by utilizing a cache for peripheral devices and communication to approximate firmware states but suffers from state explosion. Charm\cite{Talebi2018} addresses the emulation problem of smartphone drivers using the forwarding approach.

These works greatly improve the ability of firmware re-hosting, but they have two main problems. First, many devices do not have a debug channel like JTAG. Second, they could not address all hardware features correctly like interrupt and DMA. These problems greatly limit the scale of firmware re-hosting in a hardware-in-the-loop way.

\subsection{Full Emulation}
These works address the problems of re-hosting in a software-only manner. The essence is to find a way to execute hardware-related codes correctly or equivalently. These works can be divided into two categories.

\vspace{0.5mm}
\noindent \textbf{Functions Re-implementing. }
The common idea of these works is to replace the problematic functions with correct or equivalent ones. \cite{Wei2018}\cite{Wei2018a} and the following work\cite{Afek2019} boots the IOS kernel successfully by manually patching the functions that are incorrectly executed by QEMU. PowerFL\cite{Goodman2019} manually and automatically identify problematic code and stub it out with function hooks. Firmadyne\cite{Chen2016} provides an instrumented kernel to run Linux applications in firmware and researchers can analyze firmware dynamically. 
Costin et al.\cite{costin2016automated} replace the kernel in firmware with a stock kernel and emulate the whole userland of the firmware using QEMU to perform extensive web-interface testing.
HALucinator replaces functions in Hardware Abstraction Layer with equal ones. Partemu\cite{Harrison2020} addressed the problem of emulation of firmware with TrustZone, and it replaced selected components in TZOSes with a model or stub that sufficiently mimics the original components to the target.

\vspace{0.5mm}
\noindent \textbf{Emulation with Symbolic Execution. } 
These works address the problems of firmware re-hosting by applying a symbolic execution technique. FIE\cite{Davidson2013} performs symbolic execution of (MSP430 16-bit) source code and relies on KLEE\cite{Cadar2008}. Symdrive\cite{Renzelmann2012} aims to discover vulnerabilities in drivers and makes device input to the driver symbolic thereby allowing execution on the complete range of device inputs. Laelaps\cite{Cao2020} infers the expected behavior of firmware via symbolic-execution-assisted peripheral emulation and generates proper inputs to steer concrete execution on the fly.

\vspace{0.5mm}
\noindent \textbf{Model-based Emulation. }
These works emulate firmware by modeling the MMIO operations of peripherals directly. Pretender\cite{Gustafson2019} records the original device's MMIO operation then models them so that it can provide the right value to firmware when executing hardware-related code that needs a response from peripherals. P$^2$IM\cite{Feng2020} abstract diverse peripherals and handles firmware I/O on the fly basing on automatically generated models.

In fact, introducing models to firmware re-hosting brings imprecision. These models behave well when used for fuzzing. However, these are not applicable when researchers want to debug firmware for developing exploits or patches under an emulated environment.

\section{Conclusion}
In this work, we explored the abstraction in modern RTOS, and propose a novel technique called static binary-level porting to solve the firmware re-hosting problems via patching firmware. We realize this technique in a tool, FirmPorter. FirmPorter identifies Anchor Functions in 4 widely-used RTOSes and locates the existing BSP and drivers in the firmware. It can automatically patch the firmware and make it run correctly in the emulator. To address the data communication problem caused by the false emulation of some peripherals in emulators, we use a technique called Driver Hacking and create a Configurable I/O Server to exchange data between the host and the emulator. 

We prove that our method has sufficient fidelity for vulnerability assessment, that is, the PoCs or exploits developed in the emulated environment can be directly applied to real devices. We evaluated 11 firmware that contains 4 RTOS with vulnerability on each, and re-host them in unmodified emulators. The result shows that our tool can re-host the firmware correctly, and the PoCs or exploits can be executed correctly both on emulated firmware and real devices.

\bibliographystyle{plain}
\bibliography{ref}

\onecolumn
\section*{Appendix}
\label{appendix}

\subsection{Boot sequence of VxWorks}

\subsection{Boot sequence of RT-Thread}

\clearpage

\subsection{Firmware Samples}
\begin{table*}[htbp]
\centering
\begin{threeparttable}[b]
\caption{Firmware for re-hosting}
\begin{tabular}{p{2.5cm}<{\centering}p{1.85cm}<{\centering}p{2.2cm}<{\centering}p{2.5cm}<{\centering}p{2cm}<{\centering}p{4.8cm}<{\raggedright}}
    \toprule
        Firmware	&Manufactor	&Processor	&Architecture	&RTOS	&Peripherals\\
    \midrule\midrule
    		RTU	&Schinerder	&AMD LX-800	&x86	&VxWorks	&Ethernet, Disk, UART, ADC, GPIO\\
			WSN-sample	&STM	&STM32F103	&Arm Cortex-M3	&RT-Thread	&RF Transceiver, UART, ADC\\
			WSN-store	&STM	&STM32L475	&Arm Cortex-M3	&RT-Thread	&RF Transceiver, UART, SD Card\\
			WSN-upload	&STM	&STM32F429	&Arm Cortex-M4	&RT-Thread	&RF Transceiver, UART, Ethernet\\
			Music Player	&STM	&STM32F401	&Arm Cortex-M4	&RT-Thread	&UART, GPIO\\
			PLC	&Schinerder &SPEAr 680 &Arm926EJ-S &VxWorks &Ethernet, SD, UART, ADC, USB, CAN, GPIO\\
			NAS	&Samsung	&S3C6410	&Arm 1176	&VxWorks	&Ethernet, Disk, UART\\
			Car	&Raspberry Pi	&BCM2835	&Arm 1176	&RT-Thread	&GPIO, Zigbee, UART\\
			Web Server	&NXP	&LPC3130	&Arm926EJ-S	&Nuttx	&Ethernet, SD Card, UART\\
			UART Server	&NXP	&LPC1700	&Arm Cortex-M3	&Zephyr	&UART\\
			TCP Echo Server	&Microchip	&PIC32MZ EF	&MIPS32 M5150	&Nuttx	&UART, Ethernet\\

    \bottomrule
\end{tabular}
\label{tbtargets}
\begin{tablenotes}
        \footnotesize
        \item[*] \textbf{WSN} means Wireless Sensor Network
      \end{tablenotes}
      \end{threeparttable}
\end{table*}

\subsection{IDAPython Scripts for repairing symbols of VxWorks-based firmware }
\lstset{language=Python}
\begin{lstlisting}
from idaapi import *  
from idc import *  

eaStart = 0x24EF16C
eaEnd = 0x253DD54 
ea = eaStart  
eaEnd = eaEnd  
while ea < eaEnd:  
    create_strlit(Dword(ea), BADADDR)  
    sName = get_strlit_contents(Dword(ea))
    print sName
    if sName:  
        eaFunc = Dword(ea + 4)  
        MakeName(eaFunc, sName)  
        MakeCode(eaFunc)  
        MakeFunction(eaFunc, BADADDR)  
    ea = ea + 20
\end{lstlisting}

\end{document}